\documentclass[journal=jacsat,manuscript=article]{achemso}

\usepackage{chemformula} 
\usepackage[T1]{fontenc} 
\usepackage{color}
\definecolor{darkgreen}{rgb}{0,0.5,0} 

\author{J.-V. Vidal Mart\'inez-Pons}
\email{juan.vidal@uam.es}
\affiliation{Depto. de F\'isica de Materiales, Universidad Aut\'onoma de Madrid, 28049 Madrid, Spain.
}%
\alsoaffiliation{Instituto Nicol\'as Cabrera, Universidad Aut\'onoma de Madrid, 28049 Madrid, Spain.
}%

\author{S.-K. Kim}
\affiliation{Walter Schottky Institut, Institute for Advanced Study, TUM School of Computation, Information and Technology, and MCQST, Technische Universit\"at M\"unchen, 85748 Garching, Germany.}%

\author{M. Behrens}
\affiliation{Depto. de F\'isica de Materiales, Universidad Aut\'onoma de Madrid, 28049 Madrid, Spain.
}%

\author{A. Izquierdo-Molina}
\affiliation{Depto. de F\'isica de Materiales, Universidad Aut\'onoma de Madrid, 28049 Madrid, Spain.
}%

\author{A. Menendez Rua}
\affiliation{Depto. de F\'isica de Materiales, Universidad Aut\'onoma de Madrid, 28049 Madrid, Spain.
}%
\author{S. Pa{\c{c}}al}
\affiliation{Department of Physics, Izmir Institute of Technology, Izmir 35430, Turkey.
}%
\author{S. Ate{\c{s}}}
\affiliation{Faculty of Engineering and Natural Sciences, Sabanci University, 34956, Tuzla, Istanbul, Turkey.
}%

\author{L. Vi\~na}
\affiliation{Depto. de F\'isica de Materiales, Universidad Aut\'onoma de Madrid, 28049 Madrid, Spain.
}%
\alsoaffiliation{Instituto Nicol\'as Cabrera, Universidad Aut\'onoma de Madrid, 28049 Madrid, Spain.
}%
\alsoaffiliation{Instituto de F\'isica de la Materia Condensada (IFIMAC), Universidad Aut\'onoma de Madrid, 28049 Madrid, Spain.
}%

\author{C. Ant\'on-Solanas}
\email{carlos.anton@uam.es}
\affiliation{Depto. de F\'isica de Materiales, Universidad Aut\'onoma de Madrid, 28049 Madrid, Spain.
}%
\alsoaffiliation{Instituto Nicol\'as Cabrera, Universidad Aut\'onoma de Madrid, 28049 Madrid, Spain.
}%
\alsoaffiliation{Instituto de F\'isica de la Materia Condensada (IFIMAC), Universidad Aut\'onoma de Madrid, 28049 Madrid, Spain.
}%

\title
  {Temporal coherence of single photons emitted by hexagonal Boron Nitride defects at room temperature}

\keywords{hBN defects,
Single-photon emitters,
Quantum optics,
Temporal coherence,
Phonon dephasing,
Michelson interferometry}

\begin{document}

\begin{tocentry}

    \includegraphics[width=1\linewidth]{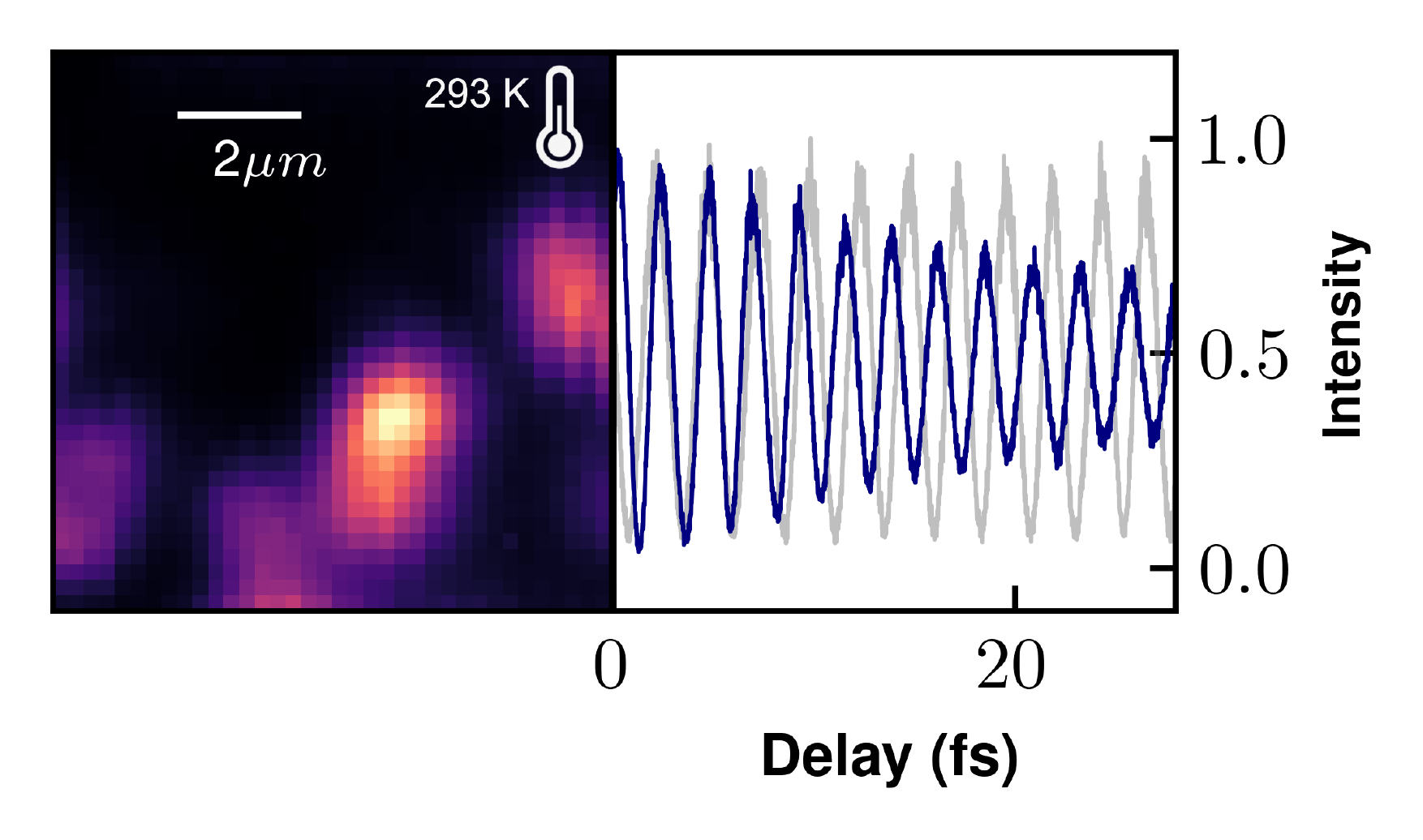}

\end{tocentry}

\begin{abstract}
 Color centers in hexagonal boron nitride (hBN) emerge as promising quantum light sources at room temperature, with potential applications in quantum communications, among others. The temporal coherence of emitted photons (i.e. their capacity to interfere and distribute photonic entanglement) is essential for many of these applications. Hence, it is crucial to study and determine the temporal coherence of this emission under different experimental conditions. In this work, we report the coherence time of the single photons emitted by an hBN defect in a nanocrystal at room temperature, measured via Michelson interferometry. The visibility of this interference vanishes when the temporal delay between the interferometer arms is a few hundred femtoseconds, highlighting that the phonon dephasing processes are four orders of magnitude faster than the spontaneous decay time of the emitter. We also analyze the single photon characteristics of the emission via correlation measurements,  defect blinking dynamics, and its Debye-Waller factor. Our room temperature results highlight the presence of a strong electron-phonon coupling, suggesting the need to work at cryogenic temperatures to enable quantum photonic applications based on photon interference.
\end{abstract}

\section{Keywords}
hBN defects,
Single-photon emitters,
Quantum optics,
Temporal coherence,
Phonon dephasing,
Michelson interferometry.

\section{Introduction}
Quantum optical technologies, such as communication, computation or metrology, demand the development of optimal quantum light sources. The two crucial characteristics of an optimal single photon source are its efficiency to generate a single photon per excitation drive, and its temporal coherence, determining its capacity to interfere and distribute entanglement \cite{senellart_high-performance_2017}. Considering solid-state sources, the state-of-the-art performance is achieved by self-assembled semiconductor quantum dots (QDs) weakly coupled to optical cavities \cite{tomm_bright_2021,ding2025high}. These results have shown record source-to-detector efficiency of $B_d{>}55$\%; in natural atoms coupled to cavities, this value is ${<}45$\% \cite{thomas_efficient_2022}. Prominent solid-state emitters, among many others \cite{aharonovich_solid-state_2016,keni_single-photon_2025}, are nitrogen- and silicon-vacancy centers \cite{fabre_quantum_2017,lukin_integrated_2020}, demonstrating fundamental applications in sensing and communications, and organic molecules \cite{toninelli_single_2021}, which present a promising route to implement multi-emitter systems via engineered dipole coupling \cite{trebbia_tailoring_2022,lange_superradiant_2024}. Over the last decade, other solid-state emitters have gained relevance, such as QDs in monolayers of transition metal dichalcogenides \cite{srivastava_optically_2015,he_single_2015,koperski_single_2015,chakraborty_voltage-controlled_2015,tonndorf_single-photon_2015}, and defects in hBN \cite{tran_quantum_2016}. Our studies along this work are based on such single photon emitter. 

Two key parameters of single photon emission performance are the intrinsic quantum efficiency of the source (ratio of the radiative spontaneous decay rate to the total decay rate) and the Debye-Waller (DW) factor (ratio of photons emitted in the zero phonon line (ZPL) to the overall spectrum, exchanging energy with phonons). The single photon lifetime (typically in the nanosecond scale and dependent on the transition dipole moment of the excited state) determines the rate at which the emitter is able to generate photons or process entanglement protocols (a photonic cavity could accelerate these timescales via the Purcell effect).

Aiming towards cryogenic-free applications, in this work, we study the single photon emission from defects in hBN nanocrystals at room temperature, extracting their two main dephasing mechanisms: the total spontaneous decay rate ($\gamma/2\pi{=}1/T_1$) and the pure dephasing rate ($\gamma^*/2\pi{=}1/T^*_2$, obtained in this work via Michelson interferometry). These two mechanisms contribute to the total dephasing rate and spectral linewidth of the emitter $\Gamma{=}\gamma{+}2\gamma^*$, where  $\Gamma/2\pi{=}2/T_2$ is the full width at half maximum (FWHM) of the ZPL \cite{wein_modelling_2021}.

The value of $\Gamma$ in certain hBN defect species has been studied via resonant spectroscopy of the ZPL as a function of temperature, showing a phonon broadening that scales as  $\Gamma \sim T^3$ \cite{sontheimer_photodynamics_2017,white_phonon_2021,horder_coherence_2022}. Other defect species in hBN nanocrystals, similar to those studied in this work, display $\Gamma \sim T^5$ \cite{ari2025temperature}. Fourier transform limited linewidths ($\Gamma{\sim}\gamma$ up to the 10 ms timescale) of certain hBN defects have been reported at room temperature \cite{dietrich_solid-state_2020,hoese_mechanical_2020}. Later theory work on Density Functional Theory simulations for $C_2 C_N$ and $V_N N_B$ defects confirm no decoupling effects from the phonon bath \cite{sharman_dft_2023}.

At cryogenic temperatures, Fourier transform-limited emission is achievable for resonant ZPL scans within ${<}10\, \mu$s, following on recent experiments with blue emitters (B-centers at 436 nm) \cite{fournier_investigating_2023,gerard_crossover_2025}. For longer timescales, the temporal coherence is limited by inhomogeneous broadening arising from spectral diffusion. This effect is caused by fluctuations of the charge distribution around the environment of the emitter, and it is typically observed in solid-state color centers\cite{white_phonon_2021}. Spectral diffusion in hBN at cryogenic temperature has been studied via photon-correlation Fourier spectroscopy \cite{brokmann2006photon}, revealing the appearance of inhomogeneous broadening at different timescales ($\sim$1 $\mu$s in Ref.\cite{sontheimer_photodynamics_2017} and $\sim$50 ns as well as longer timescales in Ref.\cite{spokoyny2020effect} ). Other techniques, such as femtosecond pump-probe spectroscopy have revealed that inhomogeneous broadening due to spectral fluctuations arises at scales as fast as $\sim$19 ps at cryogenic temperatures\cite{preuss2022resonant}. The distinction between slow jitter components, such as the inhomogeneous broadening time and slow spectral jumps could be achieved via non-linear spectroscopy techniques, such as four-wave mixing \cite{fras2016multi}. Although some factors such as the sample preparation or the used substrate have proven to play an important role on inhomogeneous broadening, so far, no significant temperature dependence has been observed on this effect\cite{akbari2021temperature}. Two-photon coalescence via Hong-Ou-Mandel interference, and off-resonant driving, has been reported for B-centers \cite{fournier_two-photon_2023}, determining a (temporally filtered) pure dephasing rate of $\gamma^*{\sim}0.8\gamma$ for consecutively emitted photons with a delay of 12.5 ns. Under resonant driving, recient studies on B-centers have measured this two-photon interference, reporting in this case an indistinguishability value of 0.92 for the same delay between consecutive single photons. \cite{gerard2025resonance} Temporal coherence of the single photon emission, under non-resonant excitation, is also characterized via Michelson interferometry. Following this method, and at cryogenic temperatures, the ZPL of hBN defects reveals a $\gamma^*{\sim}60\gamma$ \cite{sontheimer_photodynamics_2017}.

Following this trend of results, our experiments, all implemented at room temperature, investigate the temporal coherence of hBN emitters in nanocrystals. In the first part of the work, we describe the fundamental emission properties of an hBN defect (spectrum, decay dynamics, and degree of photon antibunching). In the final part of the work, our experiments determine the pure dephasing timescale of this emitter and determine their dependence as a function of the wavelength of the excitation laser. We observe a phonon-induced pure dephasing time several orders of magnitude faster than the spontaneous emission lifetime and not a significant dependence on the laser energy. 

\section{Methods}

To prepare the sample, a commercial solution of hBN nanocrystals (Graphene Supermarket, with $H_20$ as solvent) is drop-casted on a commercial distributed Bragg reflector (DBR). This is done without prior ultrasonic bath or post-annealing process. The DBR mirror consists of 10 pairs of $\text{SiO}_2$/Ti$\text{O}_2$ layers with its stopband centred at 650 nm (1.907 eV). This substrate is used to enhance the collection efficiency for the studied spectral window. In future experiments, the DBR would be part of a Fabry-Pérot cavity for the study of cavity effects in the coherence properties of the emitters \cite{grange2017reducing,mitryakhin2024engineering}. The hBN nanometric crystals are randomly scattered over the sample, with ZPLs emitting in a wide energy range, between 560 and 750 nm (1.653-2.214 eV). We reconstruct the sample topography with scanning microscopic images to nano-metrically locate defects at specific positions on the sample. The sample is navigated with XYZ closed-loop piezo-motors suitable for working at room temperature.

We implement micro-photoluminescence (PL) experiments under non-resonant laser excitation (a 450 nm Q-switch laser operated in continuous wave (CW) or pulsed regime) in a home-built confocal microscope (see setup details in the Supporting Information). In the last part of our results, we also show experinemts on several defects driven with not only 450 nm, but also 532 and 640 nm lasers. The single photon emission is collected using a 0.55 numerical aperture objective. In the collection path, the excitation laser is removed by a set of (tunable short- and long-pass) spectral filters. Then, the emission is sent to a spectrometer, or coupled into a single-mode fiber to perform time-resolved photoluminescence, Michelson interferometry ($g^{(1)}(\tau)$) or Hanbury-Brown \& Twiss correlation experiments ($g^{(2)}(\tau)$), with different sets of fiber-coupled avalanche photo-detectors ($\sim200/40$ ps jitter time and high/low detection efficiencies, respectively). 

In the Michelson interferometer, the mirror in the delay arm is attached to a piezoelectric actuator with a range of motion of 20 $\mu$m (corresponding to a fine tuneable delay of $\sim 133$ fs). In addition, this assembly is mounted on a motorized translation stage with micrometric precision, which allows us to reach longer delays in the order of tens of picoseconds (maximum spatial displacement of 5 mm, ranging from -20 to 14 ps around zero delay, according to the relative positioning of the fixed and movable mirrors). The intensity resulting from the single-photon interference in the Michelson output is re-coupled to a single-mode fiber and its count-rate is measured in a single photon detector versus temporal delay. A more thorough description of the Michelson set up is provided in the Supporting Information. The single-photon detection events (lifetime, and correlation measurements) are processed with the Extensible Time-tag Analyzer software tool \cite{lin_efficient_2021}. 

\section{Results}

\begin{figure}[h!]
    \centering
    \includegraphics[width=0.5\columnwidth]{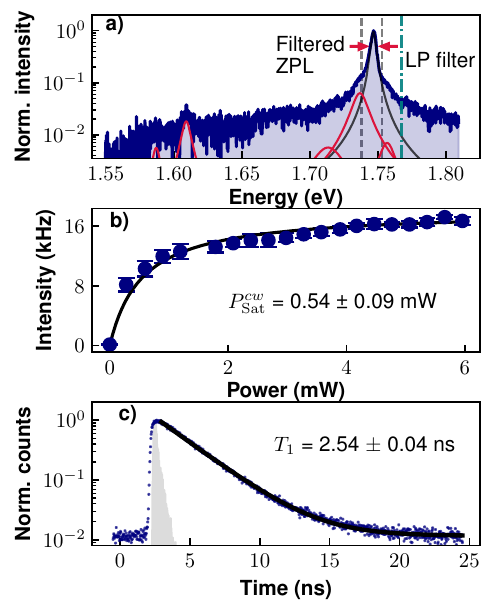}
    \caption{\textbf{Spectral and temporal characterization of the hBN emitter at room temperature}. (a) PL spectrum in log-scale under CW, non-resonant (450 nm) laser excitation, and $1.85 P^{\text{cw}}_\textrm{Sat}$ excitation power. The ZPL (black Lorentian fit) is located at $1.747$ eV, while the rest of the emission (red Lorentzian peaks) comes from the PSB. The red arrows indicate the hBN defect spectrum FWHM. The vertical dashed lines indicate the filtered spectrum of the ZPL (grey) and the full spectrum (turquoise low-energy band pass) subsequently analyzed in the Michelson interferometer. (b) ZPL pump power dependence, recording the intensity with a single-photon detector, $P^{\text{cw}}_\textrm{Sat}= 0.54$ mW. (c) Spontaneous decay of the emitter showing a mono-exponential decay $T_1 = 2.54 \pm 0.04$ ns, measured under a pump power of $1.2 P^{p}_\textrm{Sat}$. The instrument response function of the detector is included in a gray-shaded area.}
    \label{fig:Characterisation}
\end{figure}

First, we study the PL spectrum of a single hBN emitter under non-resonant CW excitation in Fig. \ref{fig:Characterisation}(a). The ZPL is identified at 1.746 eV, presenting a FWHM of 5 meV. Two asymmetric shoulders surround the ZPL peak, these low-energy (LE) absorption and emission phonon modes correspond to longitudinal acoustic phonons and a localized vibrational mode\cite{Cusco_2016,Vuong_Gil_2017,Jin_Wang_2017}. Although the exact structure of this defect is not conclusively determined, the ZPL emission could originate from different types of defects such as B-antisite, B-interstitial, or carbon substitutions in B/N vacancies\cite{wigger2019phonon, islam2024large, ari2025temperature}. To distinguish the different spectral contributions present in the phonon sideband (PSB), we fit the experimental data to the sum of several Lorentzian functions (see Fig. \ref{fig:Characterisation}(a)), accounting for the emission coming from the ZPL, the longitudinal optical (LO) modes and the LE phonon modes. We obtain a DW factor of 0.77±0.02 from the spectral fitting, and a LO contribution (to total emission) of 3\%. These values agree with previous works providing an exhaustive analysis of the spectral properties of hBN emitters at room temperature\cite{wigger2019phonon, islam2024large}. We evaluate an average DW factor of 0.43±0.18 by studying 12 different emitters in this sample (see Supp. Inf.). We note that the DW factor of the emitter shown in Fig \ref{fig:Characterisation} may be artificially enhanced by the fact that part of the PSB contribution lies in the limit of the DBR stopband, reducing its collection efficiency versus the ZPL (see Supp. Inf.). In the second part of the work, we will study the temporal coherence of the ZPL and full spectrum regions. The low energy filtering region of the full spectrum used for this study is marked with a turquoise vertical dashed line, and the vertical dashed grey lines are used to indicate the filtered ZPL region.

We continue studying the pump power dependency of the ZPL emission under CW excitation.  In Fig. \ref{fig:Characterisation} (b), the excitation power dependent intensity of the filtered ZPL is fitted with the function $I(P) = I^{cw}_{\infty}/(1+P^{cw}_{Sat}/P)$, which models the saturation dependence of a two-level system under incoherent excitation \cite{MichlerPortalupi+2024}. The saturation power is $P^{cw}_{\text{Sat}} = 0.54 \pm 0.09$ mW. The value of $I^{cw}_\infty = 18.0 \pm 0.4$ kHz indicates that the source-to-detector efficiency ($B_d$, which includes the setup and detection inefficiency), in units of the emitter lifetime ($T_1$ see below in Fig. \ref{fig:Characterisation}(c)) is $B_d\sim I^{cw}_\infty T_1{=}0.005\%$. The same saturation curve (see Supporting Information) is measured under pulsed excitation, obtaining $I^{p}_\infty = 3.2 \pm 0.1$ kHz and $P^{p}_{\text{Sat}} = 42 \pm 3$ $\mu$W; in this case, $B_d \sim 0.008\%$. The small difference between these CW and pulsed source-to-detector efficiency values may arise from a different setup performance (fiber-coupling) for these two experiments and different emitter blinking behavior in each driving regime. 

The ZPL defect lifetime is $T_1 = 2.54 \pm 0.04$ ns, as obtained from the mono-exponential fit shown in Fig. \ref{fig:Characterisation}(c); the instrument response function of the fast photon detector is indicated in a gray filled area. From the $T_1$ value, we derive a Fourier-limited linewidth of $\Gamma_{FL}/2\pi = 62.7$ MHz, several orders of magnitude narrower than the ZPL linewidth: the ZPL is strongly broadened due to electron-phonon coupling processes \cite{ari2025temperature}. The ZPL lifetime of other defects (not shown here) display similar values in the order of a few nanoseconds . The power-dependence and lifetime measurements are recorded with a low-jitter single photon detector, with an efficiency of $<30\%$ and ${\sim}40$ ps jitter time.

\begin{figure}[h]
    \centering
    \includegraphics[width=0.5\columnwidth]{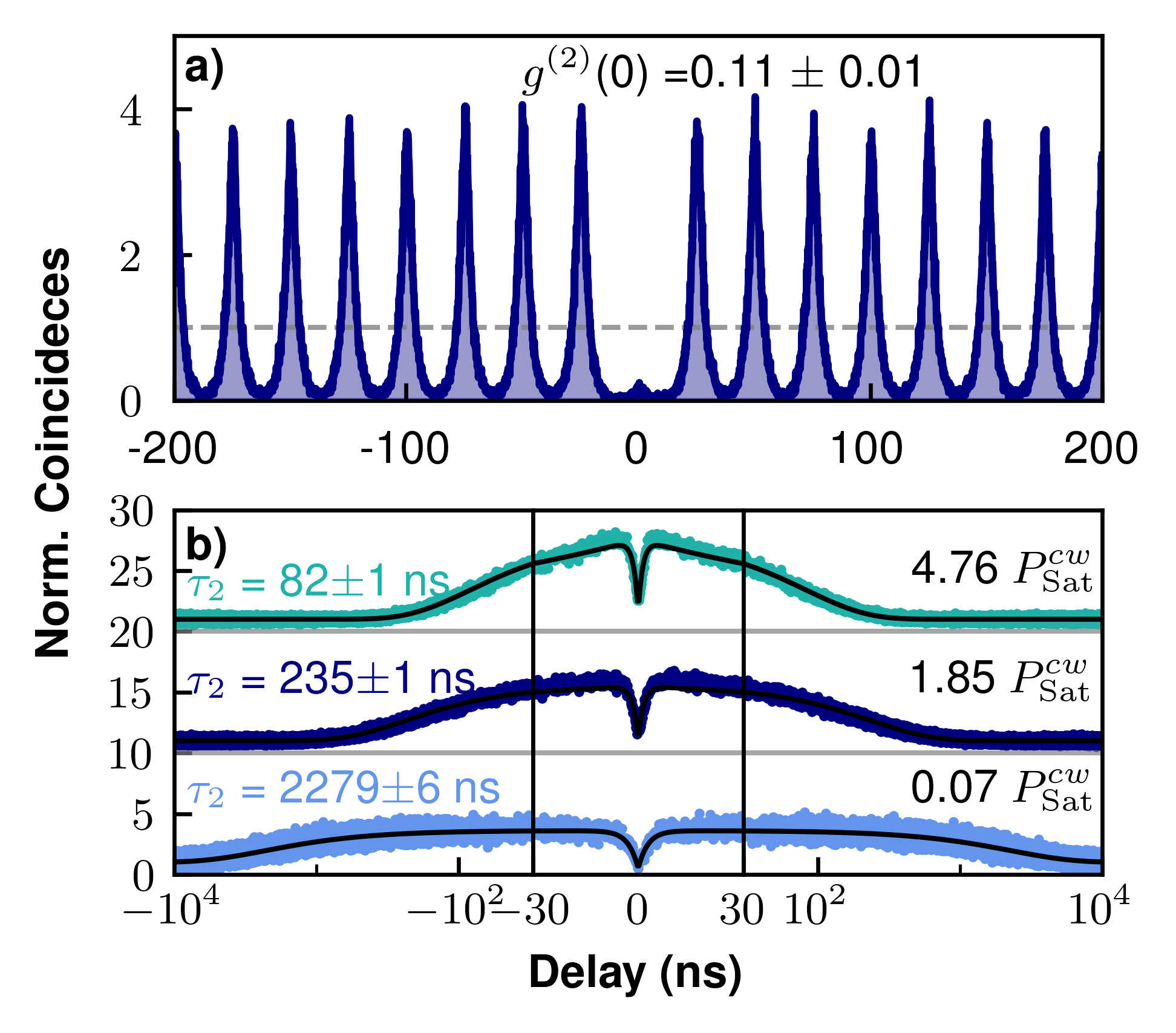}
    \caption{\textbf{Single photon character and blinking of the hBN emission}. (a) Pulsed second-order correlation function under low pump power excitation, $1.2 P^{p}_\textrm{Sat}$ laser power and 40 MHz repetition rate. The measured antibunching is $g^{(2)}(0) = 0.11\pm 0.01$ (this result does not account for the two-detector jitter). The horizontal, dashed line marks the average height of uncorrelated peaks at long delays. (b) CW second-order correlation for different pumping powers. Similarly to panel (a), the histogram normalization is done with the uncorrelated coincidence peaks at long-delays. The correlation curves are vertically displaced for clarity (the horizontal black lines at 10 and 20 normalized coincidence levels mark the correlation baseline for the medium and high driving powers). The bunching times $\tau_2$ are specified in the left side of the panel.}
    \label{fig:g2}
\end{figure}

To confirm the single photon character of the defect emission, we measure the second order correlation function $g^{(2)}(\tau)$ via a Hanbury-Brown and Twiss setup, under both pulsed and CW excitation. In pulsed regime, we obtain a value of $g^{(2)}(0) = 0.11\pm 0.01$ under $1.2 P^{p}_{\text{Sat}}$, see Fig. \ref{fig:g2}(a). Due to the emitter blinking, we note that the peaks near zero delay (not used for the $g^{(2)}(0)$ normalization) present a bunching four times more intense than the uncorrelated peaks at long delays (the gray horizontal line in this panel shows the average height of the peaks for $|\tau| {\sim} 1\, \mu s$). Although we do not discuss it here, we observe a worsening of the pulsed $g^{(2)}(0)$ value as the pulsed pump power increases, which arises from re-excitation processes during the laser pulse.

Under CW excitation and weak ($0.07 P^{cw}_{\text{Sat}}$) pump power, we measure $g^{(2)}(0) = 0.46\pm 0.13$; similarly, this value is normalized with the uncorrelated peaks at long delay and without accounting for the two-detector jitter time (${\sim} 200$ ps per detector). We note that this value is significantly larger than the $g^{(2)}(0)$ measured under pulsed excitation with higher pump power (in relation to their corresponding $P_{Sat}$). We attribute this difference in anti-bunching to the slow temporal resolution of the detectors, compared to the antibunching timescale $\tau_1$ around zero delay. This CW $g^{(2)}(\tau)$ has also been studied for $1.85 P^{cw}_{\text{Sat}}$ and $ 4.76 P^{cw}_{\text{Sat}}$ to observe the power-dependent blinking dynamics (see Fig. \ref{fig:g2}(b)). For low excitation power, there is a weak bunching effect at microsecond timescales. When pump power is increased, this timescale is reduced from 2.28 $\mu$s (0.07$P_{\text{Sat}}^{cw}$) down to  0.08 $\mu$s (4.76$P_{\text{Sat}}^{cw}$) and the bunching amplitude is increased. Such a blinking behavior is a typical signature of the presence of a dark state in a three-level ladder, affecting the emitter brightness\cite{novotny2012principles}. 

Due to re-excitation processes under CW driving, the antibunching timescale, $\tau_1$, decreases for higher pump powers. While for the lowest excitation power (0.07 $P_{\text{Sat}}^{cw}$) this value is similar to $T_1$ ($\tau^{\text{low}}_1 = 2.78 \pm 0.10$ ns), it is notably shorter for medium ($\tau^{\text{med.}}_1 = 1.49 \pm 0.04$ ns) and high ($\tau^{\text{high}}_1 = 1.31 \pm 0.03$ ns) drivings (see antibunching dips in Fig. \ref{fig:g2}(b)).  For the correlation histograms with 1.85 $P_{\text{Sat}}^{cw}$ and 4.76 $P_{\text{Sat}}^{cw}$, $\tau_1$ is close to the detectors jitter time and the blinking bunching is very prominent. In these conditions, our slow detectors can not resolve the antibunching dip and therefore the $g^{(2)}(0)$ appears overestimated.

\begin{figure}
    \centering
    \includegraphics[width=0.5\columnwidth]{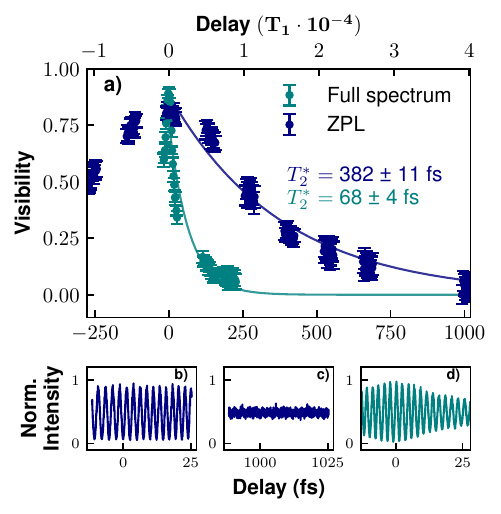}
    \caption{\textbf{Temporal coherence via Michelson interference}. (a) Fringe visibility in the output mode of the Michelson interferometer as a function of the temporal delay between the two arms for the filtered ZPL (dark blue trace) and full spectrum (turquoise). The portion of the spectrum used for each data set is indicated in Fig. \ref{fig:Characterisation} (a). Panels (b), (c) and (d) show the normalized intensity oscillations as a function of the piezo-tuned fine delay.}
    \label{fig:Visibility_filters}
\end{figure}

The mean-wavepacket overlap, i.e. indistinguishability, of the emitted single photon is determined by the ratio $\gamma/\Gamma$ (or equivalently $T_2/(2T_1)$). Apart from the ZPL resonant PL scan (which does not account for dephasing mechanisms occurring under non-resonant excitation of the defect), a precise measurement of $\Gamma$ can be obtained via Michelson interferometry \cite{santori_indistinguishable_2002,jelezko2003coherence,sontheimer_photodynamics_2017}. This measurement provides the total dephasing time of the emitted single photon along its lifetime timescale, as compared to a two-photon coalescence experiment, which captures dephasing in the timescale of the delay between two successively emitted single photons \cite{fournier_two-photon_2023}. Equation \eqref{eq:Michelsonintensity} shows the expected single-photon intensity ($N_\textrm{out}$) measured in the output interferometer arm, assuming that the emitter spectrum is Lorentzian and the two interfering modes perfectly overlap in the central beam-splitter.
\begin{align}
    N_{\textrm{out}} =\frac{1}{2}(1+e^{-\frac{\Gamma}{2}\tau} \cos (\omega_o \tau))\label{eq:Michelsonintensity}
\end{align}
In this expression, $\Gamma$ determines the exponential decay of the fringe amplitude as a function of the delay between optical paths $\tau$ ($\omega_0$ is the frequency of the Lorentzian peak). As shown in the following, our room temperature experiments set the phonon bath as the main source of decoherence, in a regime where $\gamma^* {\gg} \gamma$.

Figure \ref{fig:Visibility_filters} compiles our experiments on the temporal coherence of the single photons emitted from the defect under study. We analyze the pure dephasing rate of the filtered ZPL spectrum (dark blue symbols) and the full spectrum filtered with just a long-pass filter (turquoise data points, see filtered spectrum from Fig. \ref{fig:Characterisation}(a)). In the case of the filtered ZPL, the Lorentzian fits in Fig. \ref{fig:Characterisation}(a) show that $\sim$92\% of the intensity comes from the ZPL and $\sim$8\% from LE phonon modes, whereas contributions from other phonon modes represent less than 1\%. The pump power used for these experiments is $1.85 P^{cw}_{\text{Sat}}$. For the sake of clarity, Figs. \ref{fig:Visibility_filters}(b-d) show the interference fringes for different time delays (see timescale in the horizontal axis for each panel). Every point in panel (a) corresponds to the amplitude visibility calculated for a single oscillation period of $N_{\textrm{out}}$. 

Both sets of data (ZPL and full spectrum) are fitted with the exponential decay given in Eq. \ref{eq:Michelsonintensity}; the corresponding pure dephasing times resulting from the fits are $T_2^{*} = 382\pm 11$ fs for ZPL (dark blue trace), and $68\pm 4$ fs (turquoise trace) for the whole spectrum. In the next section we discuss the dependence of the coherence time on the filter width. We note that, at zero delay, the maximum visibility of the ZPL is $\sim 80 \%$, indicating that the spatial mode overlap of the beams interfering in the beam-splitter is not perfect. 

\begin{figure}
    \centering
    \includegraphics[width=1\columnwidth]{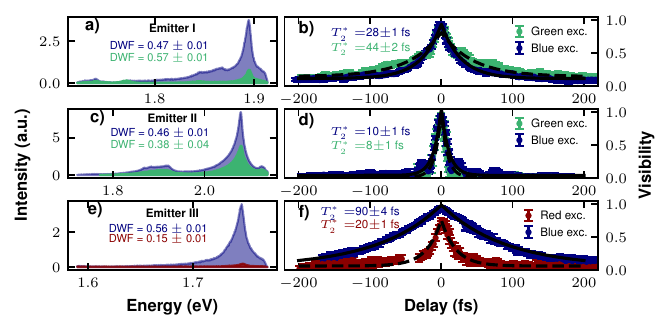}
    \caption{\textbf{Temporal coherence under different excitation energies.} Left panels (a, c, e) show the PL spectra of three different emitters (that we call emitter I, II and III respectively) driven with two excitation energies for each case. Right panels (b, d, f) display the corresponding visibility decay for the Michelson interference performed for the defects and excitation lasers on the left. Blue, green and red colors on the figure correspond with excitation energies of 2.755, 2.330 and 1.907 eV respectively. All the $T_2^*$ have been calculated from the exponential fits shown in the figures. Spectra and visibility curves have been extracted with CW excitation and power of 1 mW.}
    \label{fig:Excitation_energies}
\end{figure}

Next, we extend our measurements to study the effect of the driving energy on the coherence time of hBN single photon emission. We observe that most of the defects found with the 450 nm (blue) laser do not emit when excited with the 532 nm (green) or the 640 nm (red) laser, since there are fewer available phonon-assisted processes to drive the defect at smaller detunings. Figures \ref{fig:Excitation_energies}(a,c,e) show the PL spectra of three emitters (dubbed I, II and III, respectively, as labeled in the figure) that are successfully excited using more than one driving energy. Overall, only small differences appear when we compare the spectra, apart from a notable change in the emission intensity. Under the given experimental conditions, we observed different excitation energy dependency in the DW factor. While emitter I increases its DW factor with lower detuning excitation, emitters II and III display the opposite behavior. In the case of emitter I, the prominent LE phonon mode, present for the blue driving, disappears with green excitation. Regarding the spectral properties of emitter III, it presents a broader ZPL (FWHM  $\sim$8 meV) but very similar energy to the emitter characterized in Fig. \ref{fig:Characterisation}. It also displays a very weak LO phonon band, coherent with the description of Ref.  \cite{wigger2019phonon}. However, LE phonon modes are more important for emitter III, which makes its DW factor considerably lower, especially for the red laser excitation.

To check whether the driving energy makes a difference in the coherence properties of the emission, we perform Michelson interferometry for these defects. Figures \ref{fig:Excitation_energies} (b,d,f) show the corresponding visibility decay of emitters I, II and III (see driving conditions in the figure), respectively. Interestingly, we observe that longer $T_2^*$  correlates with higher DW factor. For instance, emitter I exhibits a longer $T_2^*$ of 44 $\pm$2 fs (green curve in Fig.\ref{fig:Excitation_energies} (b)) as the green laser excitation improves the DW factor by 0.1 compared to the blue excitation in Fig \ref{fig:Excitation_energies} (a). The change in coherence time is especially relevant for the case of emitter III. In this case, $T_2^*$ is 90$\pm$4 fs for blue excitation (see exponential fit in Figure \ref{fig:Excitation_energies} (f)), and it reduces to 20$\pm$1 fs for red laser driving. We note that in this case, the detuning of the red laser ($\sim$159 meV) lies slightly below the 1 LO phonon excitation window, which is around 165 meV\cite{wigger2019phonon}.

\section{Discussion}

There are several processes that govern the decoherence of solid-state single-photon emitters (such as electron, phonon and spin nuclei dephasing mechanisms). At low temperature, the spectral diffusion from charge fluctuations around the defect environment is the main factor for dephasing in hBN defects, as confirmed by the resonant ZPL excitation experiments of Refs. \cite{white_phonon_2021,horder_coherence_2022,fournier_investigating_2023,gerard_crossover_2025}. This inhomogeneous broadening typically increases the spectral linewidth from $\sim 60$ MHz up to $\sim 1$ GHz. Decoherence processes in hBN defects at cryogenic temperatures (8 - 80 K) have also been studied via femtosecond pump-probe spectroscopy \cite{preuss2022resonant}. Spectral jitter from homogeneous and inhomogeneous sources is identified producing an exponential and Gaussian dephasing broadening, respectively. The homogeneous dephasing time (from phonons) decreases from ~55 ps at 8K down to ~3.5 ps at 80K, with a combined emitter coherence time of ~21 ps at 8 K. In this case, the inhomogeneous dephasing arises from two-mode spectral random jumps (picosecond scale periodicity), and slower (nanosecond scale and beyond) spectral shifts. In our experiments, we probe the ultrafast timescales (below 1 ps delays, see Fig. \ref{fig:Visibility_filters}), where phonon-induced homogenous dephasing dominates over any other slower (1 ns and beyond) inhomogeneous source of decoherence, as the spectral broadening increases rapidly with $T^5$ \cite{ari2025temperature}.

Previous works at room temperature, based on a statistical description of the spectral properties of hBN defects, provide estimations for the ZPL coherence time of $\sim$100 fs for various families of emitters  \cite{wigger2019phonon}. We include two tables in the Supp. Inf. where coherence time values reported in Refs \cite{wigger2019phonon,islam2024large} are discussed. The measurement of the dephasing time via Michelson interference provides a complete picture of the emitter dephasing mechanisms (also allowing for a spectrally-selective analysis of the decoherence time); this is a particular advantage when the emitter spectrum presents a structure composed by ZPL and several phonon modes, as it is the case shown in Fig. 1(a) (see LO and LE modes), also reported in other works. \cite{marshall2011coherence, mitryakhin2024engineering,kumar2024polarization}

Coherence properties of emitters in different materials have been characterized in other works, typically at cryogenic temperatures. NV centers in diamond reveal a value of $T_2\sim$4.9 ps for the filtered ZPL (13 fs for the full spectrum), four orders of magnitude faster than its spontaneous decay time \cite{jelezko2003coherence}. Similarly, chromium centers in diamond present coherence times of $\sim$62 ps at 1.6K, which is 3\% of the corresponding lifetime  \cite{mueller2012phonon}. In self-assembled quantum dots (QDs), coherence measurements (at cryogenic temperatures) via Michelson and photon-correlation Fourier spectroscopy show values for $T_2$  ranging from few picoseconds ($\sim$10 ps for InP QDs \cite{zwiller2004single}) up to 770 ps for an InGaAs QD \cite{reigue2017probing}. At room temperature, nickel-based color centers in diamond show a $T_2\sim$210 fs \cite{marshall2011coherence}. This result is close to our reported coherent times for hBN defects, operated under similar conditions. We include a table in the Supp. Inf. where these coherence times for different platforms are discussed.

As observed in Fig. \ref{fig:Visibility_filters}(a), the restrictive spectral filtering of the ZPL (as marked in Fig. \ref{fig:Characterisation}(a)) artificially increases the coherence time and makes the temporal shape of the visibility decay to be Gaussian like, as a direct consequence of the Wiener–Khintchine theorem. We corroborate this spectrum filtering effect by simply performing the Fourier transform of the measured filtered ZPL spectrum, retrieving a very similar dephasing time as that recorded via Michelson interferometry (see Supporting Information, Fig. S6.)

It is important to note that the inhomogeneous dephasing mechanisms are too slow to take part in the Gaussian-shape visibility decay observed in the filtered spectrum of Fig. \ref{fig:Visibility_filters}(a). The temporal coherence is lost beyond 1 ps delay due to phonon-coupling at room temperature. The inhomogeneous broadening arises in the $\sim$10 ps timescale,\cite{preuss2022resonant,sontheimer_photodynamics_2017} and even longer ($\sim$10 ns and $\sim$1 $\mu$s scales).\cite{spokoyny2020effect}

From the values of the ZPL pure dephasing time, the probability of emitting two consecutive, coherent single photons under saturation conditions is ${\sim}0.015\%$. Similarly, we can expect an upper bound (i.e. assuming $100\%$ brightness) for the probability of two-photon interference in a path-delayed Mach-Zehnder interferometer. With a delay of consecutive single photons of 25ns, this value is ${\sim} 0.0015\%$. This result indicates that the strong electron-phonon coupling requires working at cryogenic temperatures. We also note that the defect lifespan is rather short in our samples (ranging between days and a few weeks); in our case, such short lifespans may be attributed to the high energy detuning between the excitation laser and the red-detuned hBN emitter spectra under study especially under blue laser excitation (between 1.7-2 eV). Previous studies in similar defects have reported similar a behavior and have argued that power-dependent optically induced local charge fluctuations \cite{boll2020photophysics} or energy-dependent photochemical reactions \cite{shotan2016photoinduced}(particularly when these defects are close to the hBN surface) might be involved. We have also studied the ZPL Michelson visibility under different pump powers, observing almost identical dephasing times (not shown here), which allows us to discard pump power induced dephasing mechanisms versus the action of phonons under our experimental conditions.

Regarding the results for different driving energies shown in Figure \ref{fig:Excitation_energies}, we can conclude that the use of very high detunings does not generally worsen the coherence time at room temperature. In this condition, the energy of excitation can be optimized considering the brightness and purity of single photon emission as the main criteria. Emitter III shows that the blue non resonant driving results in a larger coherence time that the red color laser. This wavelength-dependent behavior may involve distinct phonon-assisted excitation pathways or local photo-charge effects. Further systematic experiments versus the excitation laser wavelength will be required to clarify its origin.

\section{\label{sec:conclusion} Conclusion}

We have characterized the pure dephasing rate of the single photon emission from hBN defects at room temperature via Michelson interferometry. This dephasing time ($T_2^*$) is of the order of a few hundred femtoseconds, four orders of magnitude faster than the spontaneous decay time ($T_1$), due to the strong electron-phonon coupling mechanisms. Consequently, such room temperature single photon emission constrains the quantum photonic application landscape to protocols where photon interference is not required, such as BB84 \cite{zeng_integrated_2022,aljuboori_quantum_2023} and B92 \cite{samaner_freespace_2022,tapsin_secure_2025} or random number generation protocols \cite{white_quantum_2021,hoese_single_2022}.

In following experiments, the single photon emission from these defects will be studied at lower temperatures to attenuate the phonon dephasing rate on the temporal coherence of the emission. Coupling the defect to a photonic cavity at cryogenic temperatures will further reduce the spontaneous decay lifetime via the Purcell effect in comparison to the environmental charge fluctuations dynamics, still present at these temperatures \cite{white_phonon_2021,horder_coherence_2022,fournier_investigating_2023,gerard_crossover_2025,ari2025temperature}. Provided the large range of energies where hBN defects are present\cite{cholsuk_hbn_2024}, we believe that a reconfigurable, open Fabry-P\'erot cavity may be a suitable architecture to expand the potential applications across the visible and near-infrared bands, producing efficient and coherent single photons\cite{drawer_monolayer-based_2023}.

\begin{acknowledgement}

We acknowledge the support from the projects from the Ministerio de Ciencia e Innovaci\'on PID2023-148061NB-I00 and PCI2024-153425, the project ULTRABRIGHT from the Fundaci\'on Ram\'on Areces and the Grant “Leonardo for researchers in Physics 2023” from Fundaci\'on BBVA. This project funded within the QuantERA II Programme that has received funding from the EU H2020 research and innovation programme under GA No 101017733. SA acknowledges the support from the Scientific and Technological Research Council of Türkiye (TÜBİTAK) under GA Nos. 118F119. CA-S acknowledges the support from the Comunidad de Madrid fund “Atracci\'on de Talento, Mod. 1”, Ref. 2020-T1/IND-19785. We acknowledge Attocube for the support with the room-temperature nanopositioning system of the sample.

\end{acknowledgement}

\section{Data Availability Statement}

The data underlying this studies are openly available in the repository e-cienciaDatos at
https://doi.org/10.21950/OSZVZO.

\begin{suppinfo}

Supporting information: Experimental setup, Sample substrate: DBR mirror transmitance, Saturation curve under pulsed driving, Spectral decomposition of some emitters, Fourier Transform of the filtered ZPL spectrum, Comparison of the coherence time of other single photon emission platforms (PDF).

\end{suppinfo}

\bibliography{references}

@article{drawer_monolayer-based_2023, 
year = {2023}, 
title = {{Monolayer-Based Single-Photon Source in a Liquid-Helium-Free Open Cavity Featuring 65\% Brightness and Quantum Coherence}}, 
author = {Drawer, Jens-Christian and Mitryakhin, Victor Nikolaevich and Shan, Hangyong and Stephan, Sven and Gittinger, Moritz and Lackner, Lukas and Han, Bo and Leibeling, Gilbert and Eilenberger, Falk and Banerjee, Rounak and Tongay, Sefaattin and Watanabe, Kenji and Taniguchi, Takashi and Lienau, Christoph and Silies, Martin and Anton-Solanas, Carlos and Esmann, Martin and Schneider, Christian}, 
journal = {Nano Letters}, 
issn = {1530-6984}, 
doi = {10.1021/acs.nanolett.3c02584}, 
pmid = {37688586}, 
pmcid = {PMC10540255}, 
eprint = {2302.06340}, 
pages = {8683--8689}, 
number = {18}, 
volume = {23}
}

@article{mitryakhin2024engineering,
  title={Engineering the impact of phonon dephasing on the coherence of a WSe 2 single-photon source via cavity quantum electrodynamics},
  author={Mitryakhin, Victor N and Steinhoff, Alexander and Drawer, Jens-Christian and Shan, Hangyong and Florian, Matthias and Lackner, Lukas and Han, Bo and Eilenberger, Falk and Tongay, Seth Ariel and Watanabe, Kenji},
  journal={Physical Review Letters},
  volume={132},
  number={20},
  pages={206903},
  year={2024},
  publisher={APS}
}

@article{jelezko2003coherence,
  title={Coherence length of photons from a single quantum system},
  author={Jelezko, F and Volkmer, A and Popa, I and Rebane, KK and Wrachtrup, J},
  journal={Physical Review A},
  volume={67},
  number={4},
  pages={041802},
  year={2003},
  publisher={APS}
}

@article{senellart_high-performance_2017,
year = {2017}, 
title = {{High-performance semiconductor quantum-dot single-photon sources}}, 
author = {Senellart, Pascale and Solomon, Glenn and White, Andrew}, 
journal = {Nature Nanotechnology}, 
issn = {1748-3387}, 
doi = {10.1038/nnano.2017.218}, 
pmid = {29109549}, 
pages = {1026--1039}, 
number = {11}, 
volume = {12}
}

@article{tomm_bright_2021, 
year = {2021}, 
title = {{A bright and fast source of coherent single photons}}, 
author = {Tomm, Natasha and Javadi, Alisa and Antoniadis, Nadia Olympia and Najer, Daniel and Löbl, Matthias Christian and Korsch, Alexander Rolf and Schott, Rüdiger and Valentin, Sascha René and Wieck, Andreas Dirk and Ludwig, Arne and Warburton, Richard John}, 
journal = {Nature Nanotechnology}, 
issn = {1748-3387}, 
doi = {10.1038/s41565-020-00831-x}, 
pmid = {33510454}, 
eprint = {2007.12654}, 
pages = {399--403}, 
number = {4}, 
volume = {16}
}

@article{ding2025high,
  title={High-efficiency single-photon source above the loss-tolerant threshold for efficient linear optical quantum computing},
  author={Ding, Xing and Guo, Yong-Peng and Xu, Mo-Chi and Liu, Run-Ze and Zou, Geng-Yan and Zhao, Jun-Yi and Ge, Zhen-Xuan and Zhang, Qi-Hang and Liu, Hua-Liang and Wang, Lin-Jun and others},
  journal={Nature Photonics},
  pages={1--5},
  year={2025},
  publisher={Nature Publishing Group UK London},

}

@article{lukin_integrated_2020,
year = {2020}, 
title = {{Integrated Quantum Photonics with Silicon Carbide: Challenges and Prospects}}, 
author = {Lukin, Daniil M and Guidry, Melissa A and Vučković, Jelena}, 
journal = {PRX Quantum}, 
doi = {10.1103/prxquantum.1.020102}, 
eprint = {2010.15700}, 
pages = {020102},
number = {2}, 
volume = {1}
}

@incollection{fabre_quantum_2017,
	edition = {1},
	title = {Quantum optics with nitrogen-vacancy centres in diamond},
	isbn = {978-0-19-876860-9 978-0-19-182235-3},
	url = {https://academic.oup.com/book/27910/chapter/203958211},
	pages = {229--270},
	booktitle = {Quantum Optics and Nanophotonics},
	publisher = {Oxford University {PressOxford}},
	author = {Chu, Yiwen and Lukin, Mikhail D.},
	editor = {Fabre, Claude and Sandoghdar, Vahid and Treps, Nicolas and Cugliandolo, Leticia F.},
	urldate = {2025-04-05},
	date = {2017-05-18},
	langid = {english},
	doi = {10.1093/oso/9780198768609.003.0005},
}

@article{toninelli_single_2021,
year = {2021}, 
title = {{Single organic molecules for photonic quantum technologies}}, 
author = {Toninelli, C. and Gerhardt, I. and Clark, A. S. and Reserbat-Plantey, A. and Götzinger, S. and Ristanović, Z. and Colautti, M. and Lombardi, P. and Major, K. D. and Deperasińska, I. and Pernice, W. H. and Koppens, F. H. L. and Kozankiewicz, B. and Gourdon, A. and Sandoghdar, V. and Orrit, M.}, 
journal = {Nature Materials}, 
issn = {1476-1122}, 
doi = {10.1038/s41563-021-00987-4}, 
pmid = {33972762}, 
eprint = {2011.05059}, 
pages = {1615--1628}, 
number = {12}, 
volume = {20}
}

@article{srivastava_optically_2015, 
year = {2015}, 
title = {{Optically active quantum dots in monolayer WSe2}}, 
author = {Srivastava, Ajit and Sidler, Meinrad and Allain, Adrien V. and Lembke, Dominik S. and Kis, Andras and Imamoğlu, A.}, 
journal = {Nature Nanotechnology}, 
issn = {1748-3387}, 
doi = {10.1038/nnano.2015.60}, 
pmid = {25938570}, 
pages = {491--496}, 
number = {6}, 
volume = {10}
}

@article{he_single_2015,
year = {2015}, 
title = {{Single quantum emitters in monolayer semiconductors}}, 
author = {He, Yu-Ming and Clark, Genevieve and Schaibley, John R. and He, Yu and Chen, Ming-Cheng and Wei, Yu-Jia and Ding, Xing and Zhang, Qiang and Yao, Wang and Xu, Xiaodong and Lu, Chao-Yang and Pan, Jian-Wei}, 
journal = {Nature Nanotechnology}, 
issn = {1748-3387}, 
doi = {10.1038/nnano.2015.75}, 
pmid = {25938571}, 
pages = {497--502}, 
number = {6}, 
volume = {10}, 
}

@article{tonndorf_single-photon_2015, 
year = {2015}, 
title = {{Single-photon emission from localized excitons in an atomically thin semiconductor}}, 
author = {Tonndorf, Philipp and Schmidt, Robert and Schneider, Robert and Kern, Johannes and Buscema, Michele and Steele, Gary A. and Castellanos-Gomez, Andres and Zant, Herre S. J. van der and Vasconcellos, Steffen Michaelis de and Bratschitsch, Rudolf}, 
journal = {Optica}, 
issn = {2334-2536}, 
doi = {10.1364/optica.2.000347}, 
pages = {347--352}, 
number = {4}, 
volume = {2}
}

@article{koperski_single_2015,
year = {2015}, 
title = {{Single photon emitters in exfoliated WSe2 structures}}, 
author = {Koperski, M. and Nogajewski, K. and Arora, A. and Cherkez, V. and Mallet, P. and Veuillen, J.-Y. and Marcus, J. and Kossacki, P. and Potemski, M.}, 
journal = {Nature Nanotechnology}, 
issn = {1748-3387}, 
doi = {10.1038/nnano.2015.67}, 
pmid = {25938573}, 
pages = {503--506}, 
number = {6}, 
volume = {10}
}

@article{chakraborty_voltage-controlled_2015, 
year = {2015}, 
title = {{Voltage-controlled quantum light from an atomically thin semiconductor}}, 
author = {Chakraborty, Chitraleema and Kinnischtzke, Laura and Goodfellow, Kenneth M. and Beams, Ryan and Vamivakas, A. Nick}, 
journal = {Nature Nanotechnology}, 
issn = {1748-3387}, 
doi = {10.1038/nnano.2015.79}, 
pmid = {25938569}, 
pages = {507--511}, 
number = {6}, 
volume = {10}
}

@article{tran_quantum_2016,
year = {2016}, 
title = {{Quantum emission from hexagonal boron nitride monolayers}}, 
author = {Tran, Toan Trong and Bray, Kerem and Ford, Michael J. and Toth, Milos and Aharonovich, Igor}, 
journal = {Nature Nanotechnology}, 
issn = {1748-3387}, 
doi = {10.1038/nnano.2015.242}, 
pmid = {26501751}, 
pages = {37--41}, 
number = {1}, 
volume = {11}, 
}

@article{thomas_efficient_2022, 
year = {2022}, 
title = {{Efficient generation of entangled multiphoton graph states from a single atom}}, 
author = {Thomas, Philip and Ruscio, Leonardo and Morin, Olivier and Rempe, Gerhard}, 
journal = {Nature}, 
issn = {0028-0836}, 
doi = {10.1038/s41586-022-04987-5}, 
pmid = {36002484}, 
pmcid = {PMC9402438}, 
eprint = {2205.12736}, 
pages = {677--681}, 
number = {7924}, 
volume = {608}
}

@article{lange_superradiant_2024,
year = {2024}, 
title = {{Superradiant and subradiant states in lifetime-limited organic molecules through laser-induced tuning}}, 
author = {Lange, Christian M. and Daggett, Emma and Walther, Valentin and Huang, Libai and Hood, Jonathan D.}, 
journal = {Nature Physics}, 
issn = {1745-2473}, 
doi = {10.1038/s41567-024-02404-4}, 
eprint = {2308.08037}, 
pages = {836--842}, 
number = {5}, 
volume = {20}
}

@article{trebbia_tailoring_2022,
year = {2022}, 
title = {{Tailoring the superradiant and subradiant nature of two coherently coupled quantum emitters}}, 
author = {Trebbia, J.-B. and Deplano, Q. and Tamarat, P. and Lounis, B.}, 
journal = {Nature Communications}, 
doi = {10.1038/s41467-022-30672-2}, 
pmid = {35618729}, 
pmcid = {PMC9135760}, 
eprint = {2109.10584}, 
pages = {2962}, 
number = {1}, 
volume = {13}
}

@article{wein_modelling_2021,
year = {2021}, 
title = {{Modelling Markovian light-matter interactions for quantum optical devices in the solid state}}, 
author = {Wein, Stephen C}, 
publisher = {arXiv}, 
doi = {10.48550/arxiv.2105.06580}, 
eprint = {2105.06580}, 
note = {arXiv (quant-ph), 2105.06580, (accessed 2025-12-02)}
}

@article{white_phonon_2021, 
year = {2021}, 
title = {{Phonon dephasing and spectral diffusion of quantum emitters in hexagonal boron nitride}}, 
author = {White, Simon and Stewart, Connor and Solntsev, Alexander S and Li, Chi and Toth, Milos and Kianinia, Mehran and Aharonovich, Igor}, 
journal = {Optica}, 
doi = {10.1364/optica.431262}, 
pages = {1153}, 
number = {9}, 
volume = {8}
}

@article{horder_coherence_2022, 
year = {2022}, 
title = {{Coherence Properties of Electron-Beam-Activated Emitters in Hexagonal Boron Nitride Under Resonant Excitation}}, 
author = {Horder, Jake and White, Simon J.U. and Gale, Angus and Li, Chi and Watanabe, Kenji and Taniguchi, Takashi and Kianinia, Mehran and Aharonovich, Igor and Toth, Milos}, 
journal = {Physical Review Applied}, 
doi = {10.1103/physrevapplied.18.064021}, 
eprint = {2205.08810}, 
pages = {064021}, 
number = {6}, 
volume = {18}, 
}

@article{sontheimer_photodynamics_2017,
year = {2017}, 
title = {{Photodynamics of quantum emitters in hexagonal boron nitride revealed by low-temperature spectroscopy}}, 
author = {Sontheimer, Bernd and Braun, Merle and Nikolay, Niko and Sadzak, Nikola and Aharonovich, Igor and Benson, Oliver}, 
journal = {Physical Review B}, 
issn = {2469-9950}, 
doi = {10.1103/physrevb.96.121202}, 
eprint = {1704.06881}, 
pages = {121202}, 
number = {12}, 
volume = {96}
}

@article{sharman_dft_2023, 
year = {2023}, 
title = {{A DFT study of electron–phonon interactions for the C2CN and VNNB defects in hexagonal boron nitride: investigating the role of the transition dipole direction}}, 
author = {Sharman, K and Golami, O and Wein, S C and Zadeh-Haghighi, H and Rocha, C G and Kubanek, A and Simon, C}, 
journal = {Journal of Physics: Condensed Matter}, 
issn = {0953-8984}, 
doi = {10.1088/1361-648x/acde2b}, 
pmid = {37311467}, 
eprint = {2207.14411}, 
pages = {385701}, 
number = {38}, 
volume = {35}
}

@article{fournier_two-photon_2023, 
year = {2023}, 
title = {{Two-Photon Interference from a Quantum Emitter in Hexagonal Boron Nitride}}, 
author = {Fournier, Clarisse and Roux, Sébastien and Watanabe, Kenji and Taniguchi, Takashi and Buil, Stéphanie and Barjon, Julien and Hermier, Jean-Pierre and Delteil, Aymeric}, 
journal = {Physical Review Applied}, 
doi = {10.1103/physrevapplied.19.l041003}, 
eprint = {2210.05590}, 
pages = {L041003}, 
number = {4}, 
volume = {19}, 
}

@article{ari2025temperature, 
year = {2025}, 
title = {{Temperature-Dependent Spectral Properties of Hexagonal Boron Nitride Color Centers}}, 
author = {Ar{\i}, Ozan and Polat, Nahit and F{\i}rat, Volkan and \c{C}ak{\i}r, \"{O}zg\"u{u}r and Ate\c{s}, Serkan}, 
journal = {ACS Photonics}, 
issn = {2330-4022}, 
doi = {10.1021/acsphotonics.4c02616}, 
pages = {1676--1682}, 
number = {3}, 
volume = {12}
}

@article{lin_efficient_2021,
	title = {Efficient and versatile toolbox for analysis of time-tagged measurements},
	volume = {16},
	issn = {1748-0221},
	url = {https://iopscience.iop.org/article/10.1088/1748-0221/16/08/T08016},
	doi = {10.1088/1748-0221/16/08/T08016},
	pages = {T08016},
	number = {8},
	journaltitle = {Journal of Instrumentation},
	shortjournal = {J. Inst.},
	author = {Lin, Zuzeng and Schweickert, Lucas and Gyger, Samuel and Zwiller, V and others},
	urldate = {2025-04-15},
	date = {2021-08-01},
	langid = {english},
}

@article{samaner_freespace_2022, 
year = {2022}, 
title = {{Free‐Space Quantum Key Distribution with Single Photons from Defects in Hexagonal Boron Nitride}}, 
author = {Samaner, {\c{C}}a{\u{g}}lar and Pa{\c{c}}al, Serkan and Mutlu, G{\"o}rkem and Uyan{\i}k, K{\i}van{\c{c}} and Ate{\c{s}}, Serkan}, 
journal = {Advanced Quantum Technologies}, 
issn = {2511-9044}, 
doi = {10.1002/qute.202200059}, 
eprint = {2204.02830}, 
pages = {2200059}, 
number = {9}, 
volume = {5}, 
}

@article{aljuboori_quantum_2023, 
year = {2023}, 
title = {{Quantum Key Distribution Using a Quantum Emitter in Hexagonal Boron Nitride}}, 
author = {Al‐Juboori, Ali and Zeng, Helen Zhi Jie and Nguyen, Minh Anh Phan and Ai, Xiaoyu and Laucht, Arne and Solntsev, Alexander and Toth, Milos and Malaney, Robert and Aharonovich, Igor}, 
journal = {Advanced Quantum Technologies}, 
issn = {2511-9044}, 
doi = {10.1002/qute.202300038}, 
pages = {2300038}, 
number = {9}, 
volume = {6}, 
}

@article{white_quantum_2021, 
year = {2021}, 
title = {{Quantum random number generation using a hexagonal boron nitride single photon emitter}}, 
author = {White, Simon J U and Klauck, Friederike and Tran, Toan Trong and Schmitt, Nora and Kianinia, Mehran and Steinfurth, Andrea and Heinrich, Matthias and Toth, Milos and Szameit, Alexander and Aharonovich, Igor and Solntsev, Alexander S}, 
journal = {Journal of Optics}, 
issn = {2040-8978}, 
doi = {10.1088/2040-8986/abccff}, 
eprint = {2001.10625}, 
pages = {01LT01}, 
number = {1}, 
volume = {23}
}

@article{zeng_integrated_2022, 
year = {2022}, 
title = {{Integrated room temperature single-photon source for quantum key distribution}}, 
author = {Zeng, Helen Zhi Jie and Ngyuen, Minh Anh Phan and Ai, Xiaoyu and Bennet, Adam and Solntsev, Alexander S and Laucht, Arne and Al-Juboori, Ali and Toth, Milos and Mildren, Richard P and Malaney, Robert and Aharonovich, Igor}, 
journal = {Optics Letters}, 
issn = {0146-9592}, 
doi = {10.1364/ol.454450}, 
pmid = {35363706}, 
pages = {1673}, 
number = {7}, 
volume = {47}
}

@article{hoese_single_2022, 
year = {2022}, 
title = {{Single photon randomness originating from the symmetric dipole emission pattern of quantum emitters}}, 
author = {Hoese, Michael and Koch, Michael K. and Breuning, Felix and Lettner, Niklas and Fehler, Konstantin G. and Kubanek, Alexander}, 
journal = {Applied Physics Letters}, 
issn = {0003-6951}, 
doi = {10.1063/5.0074946}, 
eprint = {2102.09357}, 
pages = {044001}, 
number = {4}, 
volume = {120}
}

@article{cholsuk_hbn_2024, 
year = {2024}, 
title = {{The hBN Defects Database: A Theoretical Compilation of Color Centers in Hexagonal Boron Nitride}}, 
author = {Cholsuk, Chanaprom and Zand, Ashkan and Çakan, Aslı and Vogl, Tobias}, 
journal = {The Journal of Physical Chemistry C}, 
issn = {1932-7447}, 
doi = {10.1021/acs.jpcc.4c03404}, 
eprint = {2405.12749}, 
pages = {12716--12725}, 
number = {30}, 
volume = {128}, 
}

@article{santori_indistinguishable_2002, 
year = {2002}, 
title = {{Indistinguishable photons from a single-photon device}}, 
author = {Santori, Charles and Fattal, David and Vučković, Jelena and Solomon, Glenn S. and Yamamoto, Yoshihisa}, 
journal = {Nature}, 
issn = {0028-0836}, 
doi = {10.1038/nature01086}, 
pmid = {12374958}, 
pages = {594--597}, 
number = {6907}, 
volume = {419}, 
}

@book{novotny2012principles,
  title={Principles of nano-optics},
  author={Novotny, Lukas and Hecht, Bert},
  year={2012},
  publisher={Cambridge university press}
}

@article{dietrich_solid-state_2020, 
year = {2020}, 
title = {{Solid-state single photon source with Fourier transform limited lines at room temperature}}, 
author = {Dietrich, A. and Doherty, M. W. and Aharonovich, I. and Kubanek, A.}, 
journal = {Physical Review B}, 
issn = {2469-9950}, 
doi = {10.1103/physrevb.101.081401}, 
eprint = {1903.02931}, 
pages = {081401}, 
number = {8}, 
volume = {101}
}

@article{hoese_mechanical_2020, 
year = {2020}, 
title = {{Mechanical decoupling of quantum emitters in hexagonal boron nitride from low-energy phonon modes}}, 
author = {Hoese, Michael and Reddy, Prithvi and Dietrich, Andreas and Koch, Michael K. and Fehler, Konstantin G. and Doherty, Marcus W. and Kubanek, Alexander}, 
journal = {Science Advances}, 
doi = {10.1126/sciadv.aba6038}, 
pmid = {32998895}, 
pmcid = {PMC7527221}, 
eprint = {2004.10826}, 
pages = {eaba6038}, 
number = {40}, 
volume = {6}, 
}

@article{fournier_investigating_2023,
	title = {Investigating the fast spectral diffusion of a quantum emitter in {hBN} using resonant excitation and photon correlations},
	volume = {107},
	issn = {2469-9950, 2469-9969},
	url = {https://link.aps.org/doi/10.1103/PhysRevB.107.195304},
	doi = {10.1103/PhysRevB.107.195304},
	language = {en},
	number = {19},
	urldate = {2023-09-14},
	journal = {Physical Review B},
	author = {Fournier, Clarisse and Watanabe, Kenji and Taniguchi, Takashi and Barjon, Julien and Buil, StÃ©phanie and Hermier, Jean-Pierre and Delteil, Aymeric},
	month = may,
	year = {2023},
	pages = {195304},
}

@article{gerard_crossover_2025,
	title = {Crossover from inhomogeneous to homogeneous response of a resonantly driven {hBN} quantum emitter},
	volume = {111},
	issn = {2469-9950, 2469-9969},
	url = {https://link.aps.org/doi/10.1103/PhysRevB.111.085304},
	doi = {10.1103/PhysRevB.111.085304},
	language = {en},
	number = {8},
	urldate = {2025-04-20},
	journal = {Physical Review B},
	author = {G{\'e}rard, Domitille and Buil, St{\'e}phanie and Hermier, Jean-Pierre and Delteil, Aymeric},
	month = feb,
	year = {2025},
	pages = {085304},
}

@article{Jin_Wang_2017, 
year = {2017}, 
title = {{Interlayer electron–phonon coupling in WSe2/hBN heterostructures}}, 
author = {Jin, Chenhao and Kim, Jonghwan and Suh, Joonki and Shi, Zhiwen and Chen, Bin and Fan, Xi and Kam, Matthew and Watanabe, Kenji and Taniguchi, Takashi and Tongay, Sefaattin and Zettl, Alex and Wu, Junqiao and Wang, Feng}, 
journal = {Nature Physics}, 
issn = {1745-2473}, 
doi = {10.1038/nphys3928}, 
pages = {127--131}, 
number = {2}, 
volume = {13}, 
}

@article{Cusco_2016, 
year = {2016}, 
title = {{Temperature dependence of Raman-active phonons and anharmonic interactions in layered hexagonal BN}}, 
author = {Cuscó, Ramon and Gil, Bernard and Cassabois, Guillaume and Artús, Luis}, 
journal = {Physical Review B}, 
issn = {2469-9950}, 
doi = {10.1103/physrevb.94.155435}, 
pages = {155435}, 
number = {15}, 
volume = {94}, 
}

@article{Vuong_Gil_2017, 
year = {2017}, 
title = {{Exciton-phonon interaction in the strong-coupling regime in hexagonal boron nitride}}, 
author = {Vuong, T. Q. P. and Cassabois, G. and Valvin, P. and Liu, S. and Edgar, J. H. and Gil, B.}, 
journal = {Physical Review B}, 
issn = {2469-9950}, 
doi = {10.1103/physrevb.95.201202}, 
eprint = {1705.01996}, 
pages = {201202}, 
number = {20}, 
volume = {95}, 
}

@article{tapsin_secure_2025, 
title = {Secure {Quantum} {Key} {Distribution} {Using} a {Room}-{Temperature} {Quantum} {Emitter}},
	url = {http://arxiv.org/abs/2501.13902},
	doi = {10.48550/arXiv.2501.13902},
	urldate = {2025-05-05},
	publisher = {arXiv},
    author = {Tapşın, {\"O}mer S and Ağlarcı, Furkan and Pousa, Roberto G and Oi, Daniel K L and Gündoğan, Mustafa and Ateş, Serkan},	
	month = apr,
	year = {2025},
	note = {arXiv (quant-ph), 2501.13902 (accessed 2025-12-02)},
}

@article{aharonovich_solid-state_2016,
	title = {Solid-state single-photon emitters},
	volume = {10},
	copyright = {2016 Nature Publishing Group},
	issn = {1749-4893},
	url = {https://www.nature.com/articles/nphoton.2016.186},
	doi = {10.1038/nphoton.2016.186},
	language = {en},
	number = {10},
	urldate = {2018-12-12},
	journal = {Nature Photonics},
	author = {Aharonovich, Igor and Englund, Dirk and Toth, Milos},
	month = oct,
	year = {2016},
	pages = {631--641},
}

@article{keni_single-photon_2025,
	title = {Single-photon generation: materials, techniques, and the {Rydberg} exciton frontier [{Invited}]},
	volume = {15},
	copyright = {\&\#169; 2025 Optica Publishing Group},
	issn = {2159-3930},
	shorttitle = {Single-photon generation},
	url = {https://opg.optica.org/ome/abstract.cfm?uri=ome-15-4-626},
	doi = {10.1364/OME.549582},
	language = {EN},
	number = {4},
	urldate = {2025-04-15},
	journal = {Optical Materials Express},
	author = {Keni, Arya and Barua, Kinjol and Heshami, Khabat and Javadi, Alisa and Alaeian, Hadiseh},
	month = apr,
	year = {2025},
	note = {Publisher: Optica Publishing Group},
	pages = {626--643},
}

@book{MichlerPortalupi+2024,
url = {https://doi.org/10.1515/9783110703412},
title = {Semiconductor Quantum Light Sources},
title = {Fundamentals, Technologies and Devices},
author = {Peter Michler and Simone Luca Portalupi},
publisher = {De Gruyter},
address = {Berlin, Boston},
doi = {doi:10.1515/9783110703412},
isbn = {9783110703412},
year = {2024},
lastchecked = {2025-05-10}
}

@article{preuss2022resonant,
  title={Resonant and phonon-assisted ultrafast coherent control of a single hBN color center},
  author={Preuss, Johann A and Groll, Daniel and Schmidt, Robert and Hahn, Thilo and Machnikowski, Pawe{\l} and Bratschitsch, Rudolf and Kuhn, Tilmann and Michaelis de Vasconcellos, Steffen and Wigger, Daniel},
  journal={Optica},
  volume={9},
  number={5},
  pages={522--531},
  year={2022},
  publisher={Optica Publishing Group}
}

@article{brokmann2006photon,
  title={Photon-correlation Fourier spectroscopy},
  author={Brokmann, Xavier and Bawendi, Moungi and Coolen, Laurent and Hermier, Jean-Pierre},
  journal={Optics express},
  volume={14},
  number={13},
  pages={6333--6341},
  year={2006},
  publisher={Optical Society of America}
}

@article{spokoyny2020effect,
  title={Effect of spectral diffusion on the coherence properties of a single quantum emitter in hexagonal boron nitride},
  author={Spokoyny, Boris and Utzat, Hendrik and Moon, Hyowon and Grosso, Gabriele and Englund, Dirk and Bawendi, Moungi G},
  journal={The journal of physical chemistry letters},
  volume={11},
  number={4},
  pages={1330--1335},
  year={2020},
  publisher={ACS Publications}
}

@article{kumar2024polarization,
  title={Polarization dynamics of solid-state quantum emitters},
  author={Kumar, Anand and Samaner, Caglar and Cholsuk, Chanaprom and Matthes, Tjorben and Pa{\c{c}}al, Serkan and Oyun, Yag{\i}z and Zand, Ashkan and Chapman, Robert J and Saerens, Gr{\'e}goire and Grange, Rachel},
  journal={ACS nano},
  volume={18},
  number={7},
  pages={5270--5281},
  year={2024},
  publisher={ACS Publications}
}

@article{islam2024large,
  title={Large-Scale Statistical Analysis of Defect Emission in hBN: Revealing Spectral Families and Influence of Flake Morphology},
  author={Islam, Md Samiul and Chowdhury, Rup Kumar and Barthelemy, Marie and Moczko, Loic and Hebraud, Pascal and Berciaud, Stephane and Barsella, Alberto and Fras, Francois},
  journal={ACS nano},
  volume={18},
  number={32},
  pages={20980--20989},
  year={2024},
  publisher={ACS Publications}
}

@article{wigger2019phonon,
  title={Phonon-assisted emission and absorption of individual color centers in hexagonal boron nitride},
  author={Wigger, Daniel and Schmidt, Robert and Del Pozo-Zamudio, Osvaldo and Preu{\ss}, Johann A and Tonndorf, Philipp and Schneider, Robert and Steeger, Paul and Kern, Johannes and Khodaei, Yashar and Sperling, Jaroslaw and others},
  journal={2D Materials},
  volume={6},
  number={3},
  pages={035006},
  year={2019},
  publisher={IOP Publishing}
}

@article{mueller2012phonon,
  title={Phonon-induced dephasing of chromium color centers in diamond},
  author={Mueller, Tina and Aharonovich, Igor and Wang, Zhiping and Yuan, Xiangyang and Castelletto, Stefania and Prawer, Steven and Atat{\"u}re, Mete},
  journal={Physical Review B—Condensed Matter and Materials Physics},
  volume={86},
  number={19},
  pages={195210},
  year={2012},
  publisher={APS}
}

@article{marshall2011coherence,
  title={Coherence properties of a single dipole emitter in diamond},
  author={Marshall, Graham D and Gaebel, Torsten and Matthews, Jonathan CF and Enderlein, J{\"o}rg and O’Brien, Jeremy L and Rabeau, James R},
  journal={New Journal of Physics},
  volume={13},
  number={5},
  pages={055016},
  year={2011},
  publisher={IOP Publishing}
}

@article{zwiller2004single,
  title={Single-photon Fourier spectroscopy of excitons and biexcitons in single quantum dots},
  author={Zwiller, V and Aichele, T and Benson, O},
  journal={Physical Review B},
  volume={69},
  number={16},
  pages={165307},
  year={2004},
  publisher={APS}
}

@article{reigue2017probing,
  title={Probing electron-phonon interaction through two-photon interference in resonantly driven semiconductor quantum dots},
  author={Reigue, Antoine and Iles-Smith, Jake and Lux, Fabian and Monniello, L{\'e}onard and Bernard, Mathieu and Margaillan, Florent and Lemaitre, Aristide and Martinez, Anthony and McCutcheon, Dara PS and M{\o}rk, Jesper and others},
  journal={Physical review letters},
  volume={118},
  number={23},
  pages={233602},
  year={2017},
  publisher={APS}
}

@article{boll2020photophysics,
  title={Photophysics of quantum emitters in hexagonal boron-nitride nano-flakes},
  author={Boll, Mads K and Radko, Ilya P and Huck, Alexander and Andersen, Ulrik L},
  journal={Optics Express},
  volume={28},
  number={5},
  pages={7475--7487},
  year={2020},
  publisher={OSA}
}

@article{shotan2016photoinduced,
  title={Photoinduced modification of single-photon emitters in hexagonal boron nitride},
  author={Shotan, Zav and Jayakumar, Harishankar and Considine, Christopher R and Mackoit, Mazena and Fedder, Helmut and Wrachtrup, Jörg and Alkauskas, Audrius and Doherty, Marcus W and Menon, Vinod M and Meriles, Carlos A},
  journal={Acs Photonics},
  volume={3},
  number={12},
  pages={2490--2496},
  year={2016},
  publisher={ACS Publications}
}

@article{grange2017reducing,
  title={Reducing phonon-induced decoherence in solid-state single-photon sources with cavity quantum electrodynamics},
  author={Grange, Thomas and Somaschi, Niccolo and Ant{\'o}n, Carlos and De Santis, Lorenzo and Coppola, Guillaume and Giesz, Val{\'e}rian and Lema{\^\i}tre, Aristide and Sagnes, Isabelle and Auff{\`e}ves, Alexia and Senellart, Pascale},
  journal={Physical review letters},
  volume={118},
  number={25},
  pages={253602},
  year={2017},
  publisher={APS}
}

@article{gerard2025resonance,
  title={Resonance fluorescence and indistinguishable photons from a coherently driven B centre in hBN},
  author={G{\'e}rard, Domitille and Buil, St{\'e}phanie and Watanabe, Kenji and Taniguchi, Takashi and Hermier, Jean-Pierre and Delteil, Aymeric},
  year={2025},
  note ={arXiv (physics.optics), 2506.16980v2, (accessed 2025-12-02)

}
}

@article{akbari2021temperature,
  title={Temperature-dependent spectral emission of hexagonal boron nitride quantum emitters on conductive and dielectric substrates},
  author={Akbari, Hamidreza and Lin, Wei-Hsiang and Vest, Benjamin and Jha, Pankaj K and Atwater, Harry A},
  journal={Physical Review Applied},
  volume={15},
  number={1},
  pages={014036},
  year={2021},
  publisher={APS}
}

@article{fras2016multi,
  title={Multi-wave coherent control of a solid-state single emitter},
  author={Fras, F and Mermillod, Q and Nogues, G and Hoarau, C and Schneider, C and Kamp, M and H{\"o}fling, Sven and Langbein, W and Kasprzak, J},
  journal={Nature Photonics},
  volume={10},
  number={3},
  pages={155--158},
  year={2016},
  publisher={Nature Publishing Group UK London}
}

\end{document}